\documentclass[a4paper,12pt]{article}
\usepackage{graphicx}
\usepackage[usenames]{color}
\usepackage{authblk}

\newcommand{\unit}[1]{\;\mathrm{#1}}

\title{The LHC mass limits for the $SU(2)_{L+R}$ vector resonance triplet 
of a strong extension of the Standard model}

\author[1,2]{Mikul\'{a}\v{s} Gintner\thanks{gintner@fyzika.uniza.sk}}
\author[2,3]{Josef Jur\'{a}\v{n}\thanks{josef.juran@cern.ch}}

\affil[1]{Physics Department, University of \v{Z}ilina,
Univerzitn\'{a} 1, 010 26 \v{Z}ilina, Slovakia}
\affil[2]{Institute of Experimental and Applied Physics,
Czech Technical University in Prague, Horsk\'{a} 3a/22, 128 00 Prague, Czech Republic}
\affil[3]{Institute of Physics, Silesian University in Opava,
Bezru\v{c}ovo n\'{a}m. 13, 746 01 Opava, Czech Republic}

\date{}

\begin{document}

\maketitle

\begin{abstract}
In this paper, we derive the mass exclusion limits for the hypothetical vector
resonances of a strongly interacting extension of the Standard model
using the most recent upper bounds on the cross sections for
various resonance production processes. The $SU(2)_{L+R}$ triplet of
the vector resonances under consideration is embedded into the effective Lagrangian
based on the non-linear sigma model with the $125$-GeV $SU(2)_{L+R}$ scalar singlet.
No direct interactions of the vector resonance to the SM fermions are assumed.
We find that among eleven processes considered in this paper only those 
where the vector resonances decay to
$WW$ and $WZ$ provide the mass exclusion limit. 
Depending on the values of other parameters of the model
the mass limit can be as low as $1\unit{TeV}$.
\end{abstract}

  
\section{Introduction}
Even though the LHC experiments ATLAS and CMS achieved a spectacular success by 
discovering the 125 GeV Higgs boson~\cite{125GeVBosonDiscovery} 
it was more the beginning rather than 
the end of the struggle to uncover the character of physics beyond the Standard 
Model (SM). To this moment, it has not been settled down whether new physics 
takes the form of weakly coupled supersymmetry or strongly coupled composites; 
in a sense, this problem can be squeezed into the question whether the observed 
Higgs boson is a 
fundamental field or a bound state of hypothetical new strong interactions.

If the Higgs is generated as a composite state by new strong interactions
the extension of the SM can be effectively described by higher dimensional
operators that do not decouple in the low-energy limit. Presumably, they would 
modify
the SM couplings of the Higgs boson with the heavy SM fields, such as the 
electroweak (EW)
gauge bosons and/or
the third quark generation. However, while the light SM Higgs boson can 
guarantee
unitarity of the SM to virtually arbitrary high energies, this is not true 
anymore
if the Higgs couplings become anomalous~\cite{LeeQuiggThacker77,Chanowitz85}. 
Nevertheless, the least one could require from the successful
effective description of the composite state phenomenology is that it will not 
break down
at energy below the compositeness scale. Meeting this expectation might be 
assisted with by the presence of additional new composite states which 
naturally occur in 
strongly interacting theories.

Consequently, the search for new vector (and other) resonances has its rightful 
and important place in the ATLAS and CMS collaboration's activities. While no 
discovery has 
been made yet, the direct exclusion limits constantly improve. Unfortunately, the 
vector resonance limits are strongly model and parameter dependent and the 
mass exclusion limits found in the literature cover only some of the 
interesting cases.
To the best of our knowledge they do not apply to the case considered in this paper.

In this paper, the predictions for the hypothetical neutral and charged vector resonance
production cross sections times branching ratios of their various decay channels
are calculated and compared to the most recent upper bounds on this observable obtained by
the ATLAS and CMS Collaborations.
Whenever the predictions exceed the bounds the exclusion limits
for the vector resonance masses are inferred.

The effective description of the vector resonance triplet
we work with is a rather simplistic view of what 
might be observed 
at the LHC beyond the 125 GeV Higgs boson. In this description, 
the Higgs boson is a scalar composite state 
followed in the mass hierarchy by a vector composite $SU(2)$ triplet state.
In particular, the Higgs sector of the effective Lagrangian under consideration 
is based on 
the non-linear sigma model with the 125-GeV $SU(2)_{L+R}$ scalar singlet 
complementing 
the non-linear triplet of the Nambu-Goldstone bosons. The new vector resonances 
are 
explicitly present in the form of an $SU(2)_{L+R}$ triplet. This setup fits the 
situation 
when the global $SU(2)_L\times SU(2)_R$ symmetry is broken down to 
$SU(2)_{L+R}$.

The vector triplet is introduced as a
gauge field via the hidden local symmetry approach~\cite{HLS}.
Consequently, the mass eigenstate representation of the vector resonance 
contains
the admixture of EW gauge bosons.
It results in the appearance of the mixing-generated (indirect) couplings of 
the vector triplet
with all SM fermions.
The gauge sector of this effective description is equivalent to the gauge 
sector of highly-deconstructed Higgsless model with only three 
sites~\cite{3siteHiggslessModel}.

The effective model admits the introduction of the direct coupling of the 
vector resonance 
to the SM fermions. We have suggested and thoroughly investigated the 
possibility with the direct
chiral coupling exclusive to the third quark family in our previous 
works~\cite{tBESSprd11,tBESSepjc13,tBESSepjc16}. 
The model went under the name of tBESS. 
For the parameters of this model, we have established the limits based
on the low-energy data~\cite{tBESSprd11,tBESSepjc13} as well as on the most 
recent LHC measurements~\cite{tBESSepjc16}.
We use these limits to motivate the choice of numerical values throughout this 
paper.
Nevertheless, in this paper, we will restrict our predictions to the case of no 
direct interactions.
Turning the direct interactions on makes the phenomenology of the model much 
richer. Its investigation will
become subject of our following studies.

Under the given assumptions, we calculate the model's production cross section 
considering
the Drell-Yan as well as the vector boson fusion production processes with
eleven decay channels of the vector resonances. Namely, these are $WW$, $ZW$, 
$WH$, $ZH$, $jj$, $\ell\ell$, $\ell\nu$, $\tau\tau$, $bb$, $tt$, and $tb$
channels.
The cross sections are calculated as the functions of the vector 
resonance mass $M_\rho$ in the region allowed by the validity of the
utilized approximations and for various values of $g''$ spanning
the interval allowed by the limits found in~\cite{tBESSepjc16}.

The preliminary evaluations of these cross sections for several selected
values of $g''$ and $M_\rho$ and for all listed decay channels but the $\tau\tau$ 
and $bb$ ones 
appeared as a part of our recent publication~\cite{tBESSepjc16}.
There, the comparison of the cross sections to the experimental
bounds based on the ATLAS and CMS analyses of up to about $13\;\mbox{fb}^{-1}$ 
of $13\;\mbox{TeV}$ data published mostly in the middle of 2016
were performed and the mass exclusion limits derived wherever possible.
Here, we upgrade this analysis in several aspects. 
First of all, some improvements on the calculations of the cross sections
in~\cite{tBESSepjc16} have been introduced. They concern mainly the choice of
more appropriate parton distribution and vector boson fusion luminosity functions.
The improvements were a necessary step before undertaking any more complex analysis.
Nevertheless, as we report in~\cite{Comm2017} they have not resulted in any 
unexpected discrepancies with the preliminary estimates found in~\cite{tBESSepjc16}.

Secondly, where available the predicted cross sections are compared to the full
2016 data bounds based on about $36\;\mbox{fb}^{-1}$ of integrated luminosity.
The comparison results in the most up to date exclusion mass limits for the vector resonances
under considerations. In addition, the $\tau\tau$ and $bb$ channel experimental bounds 
-- not available before -- are also considered in this paper.

In next Section we will briefly overview the phenomenology of our effective 
model. 
In Section~\ref{sec:prodXS} the 
procedure for the calculation of the production cross section will be recapitulated.
Section~\ref{sec:results} will show the results obtained for various final 
state. 
The predictions
will be compared with the most recent experimental boundaries.
Finally, Section~\ref{sec:conclusions} will present the conclusions of our 
paper.

\section{The effective Lagrangian and its phenomenology}
\label{sec:effLagrangian}

The effective Lagrangian, we use in this paper, is the model we studied 
thoroughly in \cite{tBESSprd11,tBESSepjc13,tBESSepjc16}. 
It serves as the effective description 
of the LHC phenomenology of a hypothetical strongly interacting extension of 
the SM 
where the principal manifestation of this scenario would be the existence of a 
vector resonance triplet as a bound state of a new strong interactions. The 
Lagrangian is built to respect the global 
$SU(2)_L\times SU(2)_R\times U(1)_{B-L}\times
SU(2)_{HLS}$
symmetry of which the $SU(2)_L\times U(1)_Y\times SU(2)_{HLS}$ subgroup is 
also a local symmetry. The $SU(2)_{HLS}$ symmetry is an auxiliary gauge 
symmetry invoked to accommodate the $SU(2)_{L+R}$ triplet of new vector 
resonances. Each of the gauge groups is accompanied by its gauge coupling: 
$g,g',g''$, respectively. Beside the scalar singlet representing the 125 GeV 
Higgs boson and the hypothetical vector triplet, the effective Lagrangian is 
built out of the SM fields only.

In the flavor eigenstate basis, the deviations from the SM interactions of 
the gauge bosons and the vector resonance with the Higgs boson are 
parametrized by combinations of two parameters, $a_V$ and $a_\rho$.
In the mass basis, the EW gauge boson to Higgs vertices depend predominantly
on $a_V$. To a high precision, the analogical couplings of the vector resonance 
triplet to the Higgs boson are parametrized solely by $a_\rho$. The interaction 
Lagrangian terms for this sector
along with the calculations of the LHC experimental limits for $a_V$ and 
$a_\rho$ can 
be found in~\cite{tBESSepjc16}.
  
The way the description of the vector resonance is introduced implies the 
mixing between the resonance and electroweak gauge boson fields. The mixing 
induces the (indirect) couplings between the vector resonance and fermions that 
are proportional to $1/g''$.
Even though the considered symmetry also admits the introduction of 
the direct interactions of the vector resonance with fermions we will restrict
our analysis to
the situation when the vector resonance couples to fermions only via the mixing 
induced interactions.

The masses of the charged and neutral vector resonances in the model are 
virtually degenerate. The leading order formula for the mass reads
\begin{equation}
   M_\rho = \frac{\sqrt{\alpha}g''}{2}v,
\end{equation}
where $\alpha$ is a dimensionless free parameter emerging in the effective Lagrangian
and $v$ is the electroweak symmetry breaking scale.
Usually, $\alpha$ is traded off for $M_\rho$ so that the latter can serve as 
one of the free parameters of the model.
Our previous studies of the low-energy limits~\cite{tBESSprd11,tBESSepjc13} as well as
the Higgs-related limits and the
unitarity limits~\cite{tBESSepjc16} suggest that we should consider 
$M_\rho\geq 1\;\mbox{TeV}$ and $10\leq g''\leq 20$. 

The Higgs-related parameters $a_{V}$ and $a_{\rho}$
influence significantly only decay channels
of very small branching ratios. Hence, their impact on the
cross sections calculated in this paper is very negligible.
Thus, throughout this paper, we will use $a_V=1$ (the SM case) and 
$a_\rho=0$ (no Higgs-to-vector resonance coupling). These values are 
quite close to one of the experimentally preferred points of the parameter 
space~\cite{tBESSepjc16}. 

\begin{figure}[htb]
\centerline{%
\includegraphics[width=12.5cm]{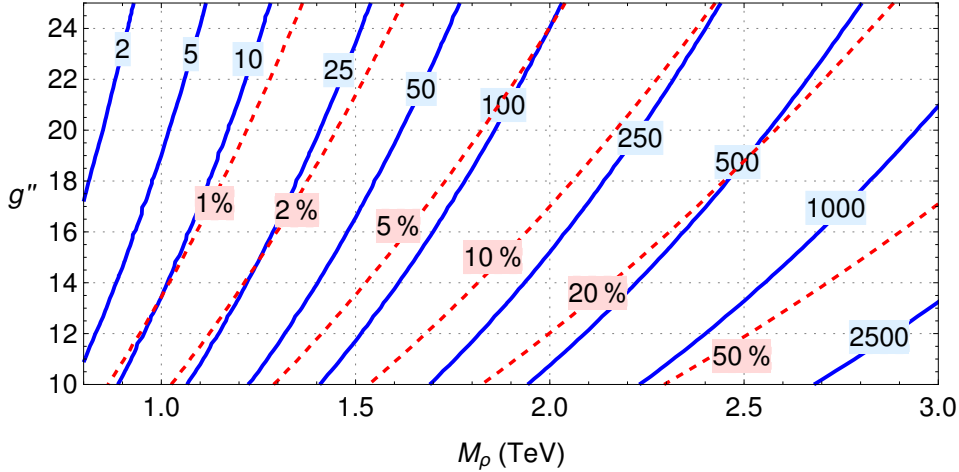}}
\caption{The solid lines represent the contours of the total decay width of the 
vector resonance (labeled in GeV) in the $g''-M_\rho$ parameter space. The 
dashed lines represent the width-to-mass ratio of the vector resonance (labeled 
in percents). No direct interactions of the resonance to fermions are assumed.}
\label{Fig:DecayWidthContours}
\end{figure}

In the case of no direct interactions of the vector resonance with fermions,
the total decay width of the resonance can be well approximated by
\begin{equation}
   \Gamma_\rho = \frac{1}{48\pi v^4}\;\frac{M_\rho^5}{g^{\prime\prime 2}}.
\end{equation}
In Fig.~\ref{Fig:DecayWidthContours} we depict how the vector resonance total width
depends on the resonance mass and $g''$. 
At the same time, the graph shows
the width-to-mass ratio contours. This information will be important
later in the paper when the region of the validity of the used approximations
is considered.

\section{The production cross section calculations}
\label{sec:prodXS}

We are interested in the LHC cross sections of the two-particle final state 
processes that would proceed via the vector resonance production. We calculate 
the cross section $\sigma(pp\rightarrow ab+X)$ 
using the Narrow Width Approximation (NWA), 
i.e. as the product of the on-shell production 
cross section of the resonance, $\sigma_\mathrm{prod}$, and the branching ratio 
for the vector resonance decay channel under consideration 
\begin{equation}\label{eq:xsNWAformula}
   \sigma(pp\rightarrow ab+X)\stackrel{\mathrm{NWA}}{=} 
   \sigma_\mathrm{prod}(pp\rightarrow \rho+X)\times\mathrm{BR}(\rho\rightarrow 
ab).
\end{equation}
As the name suggests the NWA works when $\Gamma_\rho\ll M_\rho$. 
It also ignores the signal-background 
interference effects. The influence of these effects on the precision of the 
approximation have been inspected in~\cite{Pappadopulo_etal14}.

The production cross section of a resonance can be expressed as
\begin{equation}\label{eq:ProdXS}
   \sigma_\mathrm{prod}(pp\rightarrow \rho+X)=\sum_{i\leq j\in p} 16\pi^2 K_{ij}
   \frac{\Gamma_{\rho\rightarrow 
ij}}{M_\rho}\frac{dL_{ij}}{d\hat{s}}|_{\hat{s}=M_\rho^2},
\end{equation}
where $i$, $j$ run through all partons of the colliding protons and 
$\Gamma_{\rho\rightarrow ij}$ is the partial decay width of the resonance to 
the partons $i$ and $j$. Furthermore, $dL_{ij}/d\hat{s}$ is the luminosity 
of the colliding partons, and
\begin{equation}
   K_{ij} = \frac{2J+1}{(2S_i+1)(2S_j+1)}\frac{C}{C_iC_j},
\end{equation}
where $J$ is the spin of the resonance, $C$ is its color factor, and $S_i$, 
$S_j$ and 
$C_i$, $C_j$ are the spins 
and colors of the initial partons, respectively. Note that a model dependence 
enters the production cross section (\ref{eq:ProdXS}) virtually\footnote{
In principle, the parton-parton luminosity is also sensitive to new physics via 
modifications of the SM couplings and the parton distribution functions. 
Nevertheless, we expect these effects to be negligible and ignore them in our 
analysis.
} 
only via the partial decay width $\Gamma_{\rho\rightarrow ij}$.

Two dominant production mechanisms for the triplet of our vector resonances are 
the Drell-Yan (DY)  and the vector boson fusion (VBF) processes. We will 
consider them both in our analysis. For the sake of simplicity, in our 
calculations of the production cross sections the proton contents is reduced 
down to the up and down quarks\footnote{
We do not expect that this approximation would significantly influence 
conclusions when there are no direct interactions of the vector 
resonance to fermions.
}.

In the DY case, the parton-parton luminosity in (\ref{eq:ProdXS}) is defined as
\begin{equation}
   \frac{dL_{ij}}{d\hat{s}}=\frac{1}{s}\int_\tau^1 
   \frac{dx}{x}\frac{1}{1+\delta_{ij}}
   [f_i(x,\hat{s})f_j(\tau/x,\hat{s})+i\leftrightarrow j],
\end{equation}
where $s$ and $\hat{s}$ are the squared center of mass energies of the 
colliding 
protons and partons, respectively, $\tau=\hat{s}/s$, and $f_i$ is a parton 
distribution function of the $i$th parton with the momentum fraction $x$ of its 
proton's momentum.

The VBF production will be calculated in the Effective $W$ Approximation 
(EWA)~\cite{Dawson1985EWA}.
When the EWA is applied to the VBF case, the W and Z bosons are also treated as 
partons 
in the proton.
The VBF luminosity can be expressed as
\begin{eqnarray}
   \frac{dL_{V_m V_n[pp]}}{d\tau}&=&\sum_{i\leq j} \frac{1}{1+\delta_{ij}}
   \int_\tau^1 \frac{dx_1}{x_1} \int_{\tau/x_1}^1 \frac{dx_2}{x_2}
   \nonumber\\
   && \times
   [f_i(x_1,q^2) f_j(x_2,q^2) \frac{dL_{V_mV_n[q_iq_j]}}{d\hat{\tau}}
   +i\leftrightarrow j],
\end{eqnarray}
where $\hat{\tau}=\tau/(x_1 x_2)$, and  $dL_{V_mV_n[q_iq_j]}/d\hat{\tau}$
is the luminosity for two vector bosons $V_m$ and $V_n$ emitted from $i$th and 
$j$th quarks, respectively. In the EWA, the latter can be obtained analytically
assuming that the gauge bosons are emitted on-shell and in small angles to 
their 
parental quarks. Also, if the gauge bosons fuse to a heavy resonance their 
masses should be negligibly small compare to the resonance mass. In 
this approximation, the transversal and longitudinal polarizations of the 
emitted gauge bosons are considered as separate modes.
Since the longitudinal mode usually dominates in the presence of the deviations
from the SM, we use the longitudinal mode only in our calculations.

The luminosity for two longitudinal vector bosons $V_m$ and $V_n$ emitted from 
$i$th and $j$th quarks reads
\begin{equation}
   \frac{dL_{V_m^LV_n^L[q_iq_j]}}{d\hat{\tau}} =
   \frac{v_{m[i]}^2+a_{m[i]}^2}{4\pi^2} \frac{v_{n[j]}^2+a_{n[j]}^2}{4\pi^2}
   \frac{1}{\hat{\tau}} [(1+\hat{\tau})\log(1/\hat{\tau})-2(1-\hat{\tau})],
\end{equation}
where $v_{m[i]}$ and $a_{m[i]}$ are the vector and axial couplings of the gauge 
boson $V_m$ to the quark current $q_i$. In particular,
\begin{equation}
   v_{W[q_i]} = -a_{W[q_i]} = \frac{g}{2\sqrt{2}}
\end{equation}
for any $q_i$, and
\begin{equation}
   v_{Z[u]} = \frac{g}{4c_W}(1-\frac{8}{3}s_W^2),\;\;\;
   a_{Z[u]} = \frac{g}{4c_W},
\end{equation}
\begin{equation}
   v_{Z[d]} = -\frac{g}{4c_W}(1-\frac{4}{3}s_W^2),\;\;\;
   a_{Z[d]} = -\frac{g}{4c_W},
\end{equation}
where $s_W=\sin\theta_W$ and $c_W=\cos\theta_W$.

\begin{figure}[htb]
\centerline{%
\includegraphics[width=12.5cm]{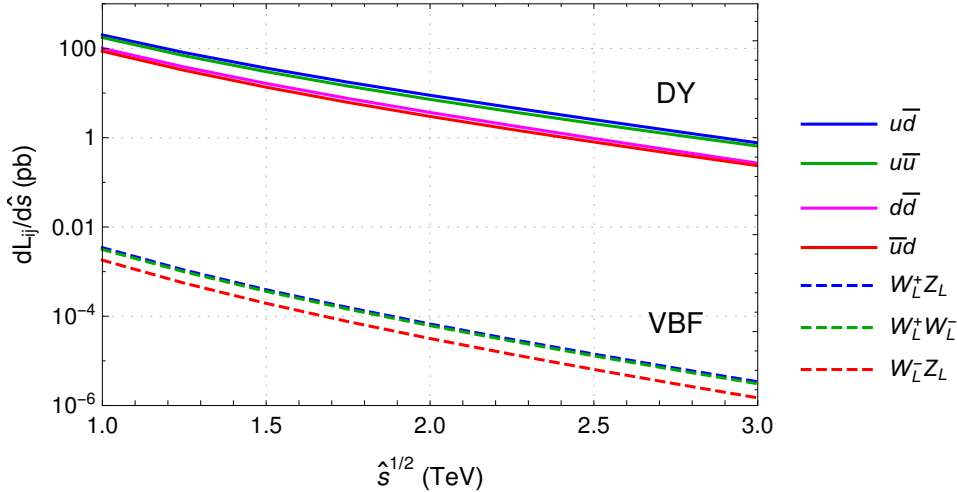}}
\caption{The parton-parton luminosities for the DY (solid lines) and 
longitudinal VBF (dashed lines) production in the proton-proton collisions 
at $\sqrt{s}=13\;\mbox{TeV}$. 
The CT10 set of parton distribution functions was used.}
\label{Fig:PartonLuminosities}
\end{figure}

For the numerical evaluation of the parton-parton luminosity  we used the 
Mathematica~\cite{Mathematica10_2016} package Mane Parse~\cite{ManeParse2016} 
with the PDF set CT10 from the LHAPDF 6 library at HepForge 
repository~\cite{HepForgeCT10_2015}. The obtained parton-parton 
luminosities for both production mechanisms of the new vector resonance at the 
LHC ($\sqrt{s}=13\;\mbox{TeV}$) 
are depicted in Fig.~\ref{Fig:PartonLuminosities}.

\section{The vector resonance mass limits and the upper bounds on
$\sigma(pp\rightarrow \rho+X)\times\mbox{BR}$}
\label{sec:results}

Once the production cross sections have been evaluated, we are a single step 
from  finalizing the predictions of the LHC cross sections of the processes 
under investigation. The final step involves the 
multiplication of the production cross section by an appropriate branching 
ratio (see Eq.~(\ref{eq:xsNWAformula})).
When the direct interactions of the vector resonances are considered,
the BR endows the resulting cross section with the sensitivity to the 
corresponding model's parameters.

We can evaluate how the existing ATLAS and CMS data restrict our model when we 
compare its predictions to the upper bounds
on the resonance production cross section times the branching ratios for various 
decay channels. The bounds are rather model independent once
spin of the resonance under consideration is specified. Of course,
one should keep in mind that the calculations involved proceed under the 
assumption of a narrow-width resonance.
As can bee seen in Fig.~\ref{Fig:DecayWidthContours}, increasing $M_\rho$
takes us away from the NWA region. Consequently, our cross section predictions
become less reliable.
On the other hand, as $M_\rho$ grows the cross sections also depart from 
the experimental bounds. Thus our predictions of the cross sections
can serve their purpose even at a higher mass to a certain extent. 
Nevertheless, we do not
consider as meaningful to go beyond $M_\rho=3\unit{TeV}$ in our analysis.

As we will see below the $WW$ and $WZ$ channels are
the only ones among those investigated in this paper in which the current data
restrict our model. The restrictions can be translated into the lower exclusion 
bounds on the vector resonance mass.

\subsection{The $WW$ and $WZ$ channels}

In the absence of the direct couplings of the vector resonance triplet to 
fermions
the decay widths of the neutral and charged vector resonances are dominated
by their decays to the EW gauge bosons: $\mathrm{BR}(\rho\rightarrow 
WW/WZ)>99\%$.

\begin{figure}[htb]
\centerline{%
\includegraphics[width=12.5cm]{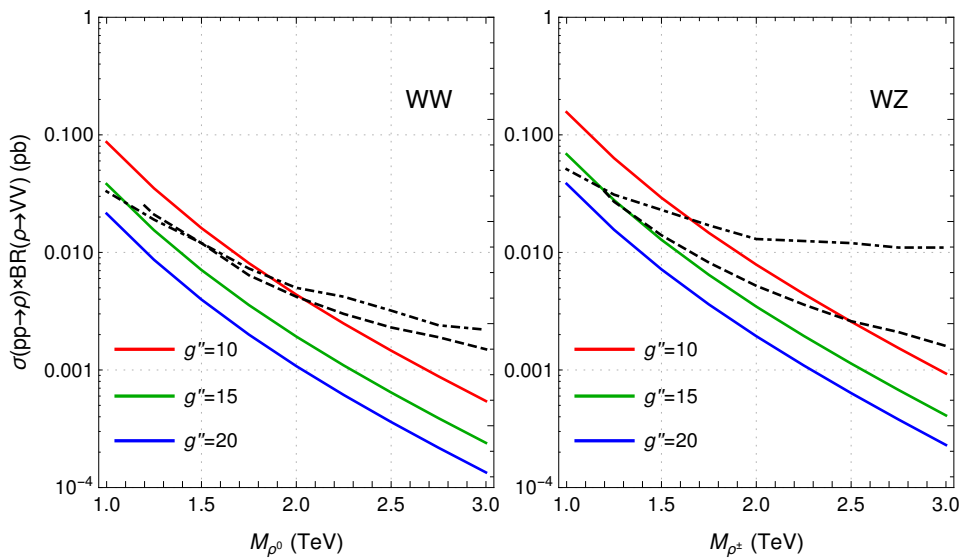}}
\caption{The $13\unit{TeV}$ cross section times branching ratios of our model
for the $WW$ (left panel) and $WZ$ (right panel) decay
channels at $g''=10$ (solid upper red), $15$ (solid middle green), and $20$ (solid bottom blue).
In both panels, the experimental $95\%$~C.L. upper limit dashed curves 
are based on the dijet final state data ($35.9\unit{fb}^{-1}$)~\cite{ExpXS-N1-B2G-17-001}.
The dot-dashed $WW$ and $WZ$ curves 
are based on the semileptonic final state data ($13.2\unit{fb}^{-1}$),
\cite{ExpXS-2016-AC062} and \cite{ExpXS-2016-AC082}, respectively.}
\label{Fig:XSxBRWW}
\end{figure}

In Fig.~\ref{Fig:XSxBRWW},
we present the cross section times branching ratios for the $WW$ and $WZ$ decay
channels of our model at the LHC collision energy of $13\unit{TeV}$.
The predictions are given for three different values of $g''$, namely $10$, 
$15$, and $20$.
The $g''$ values were chosen to span the region allowed by the combination of 
the limits obtained in~\cite{tBESSepjc16}. In addition, the most restrictive
upper $95\%$~C.L. experimental bounds on the cross section times branching ratio 
are superimposed in the graphs. In fact, since the current experimental data 
do not contradict the expectation curves we take the latter as the established
experimental boundary.

In the $WW$ channel, the experimental upper bounds~\cite{ExpXS-N1-B2G-17-001,ExpXS-2016-AC062}
exclude the mass of the neutral vector resonance below about $2.1\unit{TeV}$ and 
$1.1\unit{TeV}$ for $g''=10$ and $15$, respectively.
The $g''=20$ case is not restricted above $1\unit{TeV}$.
In the $WZ$, the experimental upper bounds~\cite{ExpXS-N1-B2G-17-001,ExpXS-2016-AC082}
exclude the mass of the charged vector resonance below about $2.5\unit{TeV}$ and 
$1.3\unit{TeV}$ for $g''=10$ and $15$, respectively. Again, the $g''=20$ case is 
not restricted above $1\unit{TeV}$. In our model, the neutral and charged
vector resonances are virtually degenerate in their masses. Therefore, within the model,
the stronger exclusion mass limit on the charged resonance can be considered as the limit
for the neutral resonance, as well.

While we do not consider the direct interactions of the vector resonance with 
fermions in this paper, we can briefly estimate the effect of the direct interaction
with the third quark generation as it was introduced in our 
tBESS model~\cite{tBESSprd11,tBESSepjc13}.
There, the corresponding chiral couplings were parameterized by the $b_{L,R}$ free parameters.
Setting $b_{L,R}$ to their maximally low-energy precision data allowed values
--- $b_{L,R}\approx 0.1$~\cite{tBESSepjc13} --- can lower
$\mathrm{BR}(WW)$ of the $1\unit{TeV}$ resonance down
to about $70\%$ for $g''=10$, to $30\%$ for $g''=15$,
and to $12\%$ for $g''=20$. In the $2\unit{TeV}$ resonance case, $\mathrm{BR}(\rho\rightarrow WW)$ 
would be lowered to about $97\%$, $87\%$, and $67\%$, respectively.
Thus, the direct fermionic interactions of the vector resonance can noticeably decrease 
the cross section predictions (and, thus, release the experimental restrictions) of 
the model in this channel. The similar effect occurs in the $ZW$ channel.
When $M_\rho=1\unit{TeV}$ and $b_{L,R}=0.1$, $\mathrm{BR}(WZ)$
gets lowered to about $71\%$, $31\%$, and $12\%$ for $g''=10$, $15$, and $20$, respectively.
For $M_\rho=2\unit{TeV}$, the corresponding BR's read
$97\%$, $87\%$, and $67\%$.

\subsection{The $bb$, $tt$, and $tb$ channels}

\begin{figure}[htb]
\centerline{%
\includegraphics[width=12.5cm]{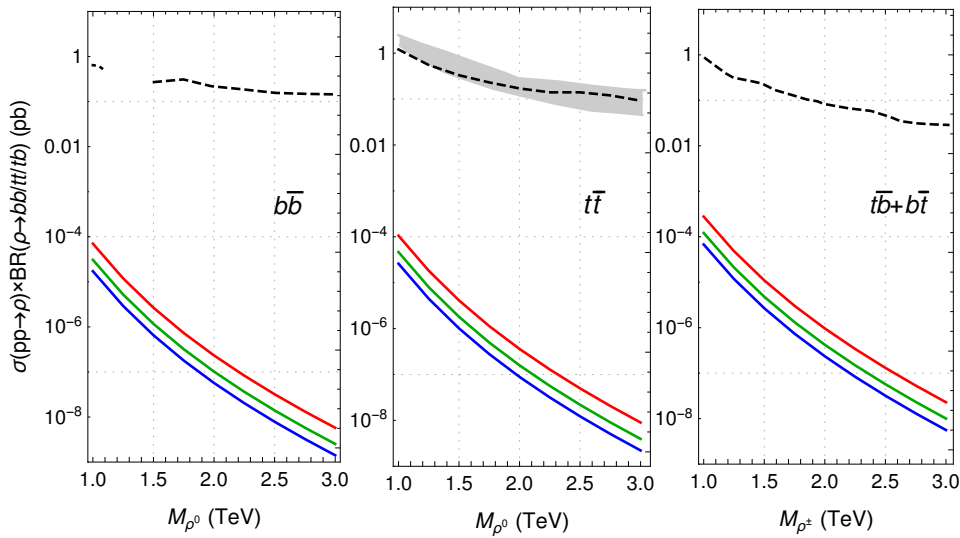}}
\caption{The $13\unit{TeV}$ cross section times branching ratios of our model
for the $bb$ (left panel), $tt$ (middle panel), and $tb$ (right panel) decay
channels at $g''=10$ (solid upper red), $15$ (solid middle green), and $20$ (solid bottom blue).
In the $bb$ channel, the experimental $95\%$~C.L. upper limit curves 
are based on $3.2\unit{fb}^{-1}$~\cite{ExpXS-2016-AC031} (dotted) 
and $13.3\unit{fb}^{-1}$ \cite{ExpXS-2016-AC060} (dashed) of data.
In the $tt$ channel, the bounds are based 
on the $2.6\unit{fb}^{-1}$ lepton+jets \& fully hadronic final 
states~\cite{ExpXS-2017-EP-2017-049} (gray stripe) 
and the $3.2\unit{fb}^{-1}$ lepton+jets final state~\cite{ExpXS-2016-AC014} (dashed).
Finally, the $tb$ channel experimental limit originate 
from the $35.9\unit{fb}^{-1}$ lepton+jets final state~\cite{ExpXS-N3-B2G-17-010} (dashed).}
\label{Fig:XSxBRbbtt}
\end{figure}

In the left panel of Fig.~\ref{Fig:XSxBRbbtt},
we present the cross section times branching ratio for the $bb$ channel
of our model at the LHC collision energy of $13\unit{TeV}$.
In the middle panel, there is the prediction for the $tt$ channel shown.
The $tb$ channel is depicted in the right panel of Fig.~\ref{Fig:XSxBRbbtt}.
There, the sum of the $t\bar{b}$ and $b\bar{t}$ contributions is considered.
The predictions are given for $g''=10$,  $15$, and $20$.
When there is no direct interaction the predicted cross sections in these channels
are not far from each other.
In the $bb$ and $tt$ channels, the existing upper experimental 
bounds~\cite{ExpXS-2016-AC031,ExpXS-2016-AC060,ExpXS-2017-EP-2017-049,ExpXS-2016-AC014} 
are several orders of magnitude above
the predicted cross sections. The same situation can be found in the $tb$ channel
(for the experimental upper bounds for this channel, 
see~\cite{ExpXS-N3-B2G-17-010}). Thus there are currently no exclusion limits
on the no-direct-interaction version of our model resulting from these channels.

However, once the direct interaction to the third quark family is turned on
the $bb$, $tt$ and $tb$ channels will be affected the most.
For example, setting $b_{L,R}=0.1$ the branching ratios for these channels increase
from $\mathrm{BR}(bb)=0.08\%(0.005\%)$, $\mathrm{BR}(tt)=0.12\%(0.008\%)$, and 
$\mathrm{BR}(tb)=0.18\%(0.012\%)$ for $M_\rho=1\unit{TeV}$ ($2\unit{TeV}$),
to the values ranging in 
$15\%(1.4\%)\leq \mathrm{BR}(bb)\leq 44\%(17\%)$,
$14\%(1.3\%)\leq \mathrm{BR}(tt)\leq 44\%(16\%)$,
and $29\%(2.7\%)\leq \mathrm{BR}(tb)\leq 87\%(33\%)$ when
$10\leq g''\leq 20$. 
Yet it does not seem to be enough
to obtain any exclusion limits from the latest measurements.

\subsection{The remaining channels}

Beside the channels discussed above we have also calculated predictions for 
the following channels: $ZH$, $WH$, $jj$, $\ell\ell$, $\ell\nu$, and $\tau\tau$,
where $\ell=e,\mu$.
\begin{figure}[htb]
\centerline{%
\includegraphics[width=12.5cm]{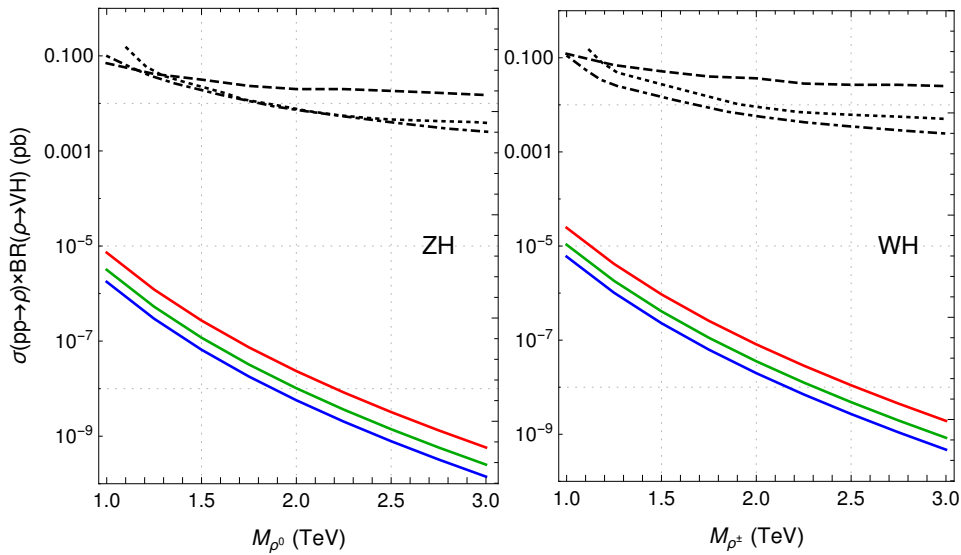}}
\caption{The $13\unit{TeV}$ cross section times branching ratios of our model
for the $ZH$ (left panel) and $WH$ (right panel) decay
channels at $g''=10$ (solid upper red), $15$ (solid middle green), and $20$ (solid bottom blue).
The experimental upper bounds shown in both panels originate
from the following measurements:
the ATLAS $36.1\unit{fb}^{-1}$ $qqbb$ final state~\cite{ExpXS-2017-AC018} (dotted),
the CMS $35.9\unit{fb}^{-1}$ $qqbb$ final state~\cite{ExpXS-2017-CP002} (dot-dashed),
and the ATLAS $3.2\unit{fb}^{-1}$ $\ell\ell bb$+$\nu\nu bb$/$\ell\nu bb$
final state~\cite{ExpXS-2016-AarXiv2} (dashed).}
\label{Fig:XSxBRVH}
\end{figure}
\begin{figure}[htb]
\centerline{%
\includegraphics[width=12.5cm]{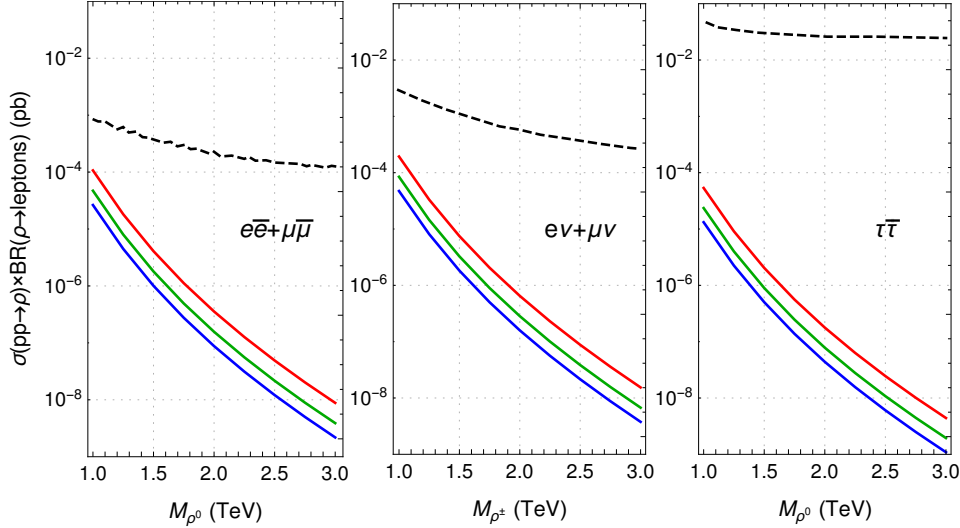}}
\caption{The $13\unit{TeV}$ cross section times branching ratios of our model
for $e^+e^-+\mu^+\mu^-$ (left panel), $e^+\nu_e+\mu^+\nu_\mu+\mathrm{c.c.}$ 
(middle panel) and $\tau^+\tau^-$ (right panel) decay
channels at $g''=10$ (solid upper red), $15$ (solid middle green), and $20$ (solid bottom blue).
The experimental upper bounds (dashed) for $e^+e^-+\mu^+\mu^-$ are based on
$36.1\unit{fb}^{-1}$ of data~\cite{ExpXS-2017-AC027}, 
on $36.1\unit{fb}^{-1}$ of data for 
$e^+\nu_e+\mu^+\nu_\mu+\mathrm{c.c.}$~\cite{ExpXS-2017-AC016}, 
and on $2.2\unit{fb}^{-1}$ of data for $\tau^+\tau^-$~\cite{ExpXS-2017-CMS-JHEP02}.}
\label{Fig:XSxBRleptons}
\end{figure}
\begin{figure}[htb]
\centerline{%
\includegraphics[width=12.5cm]{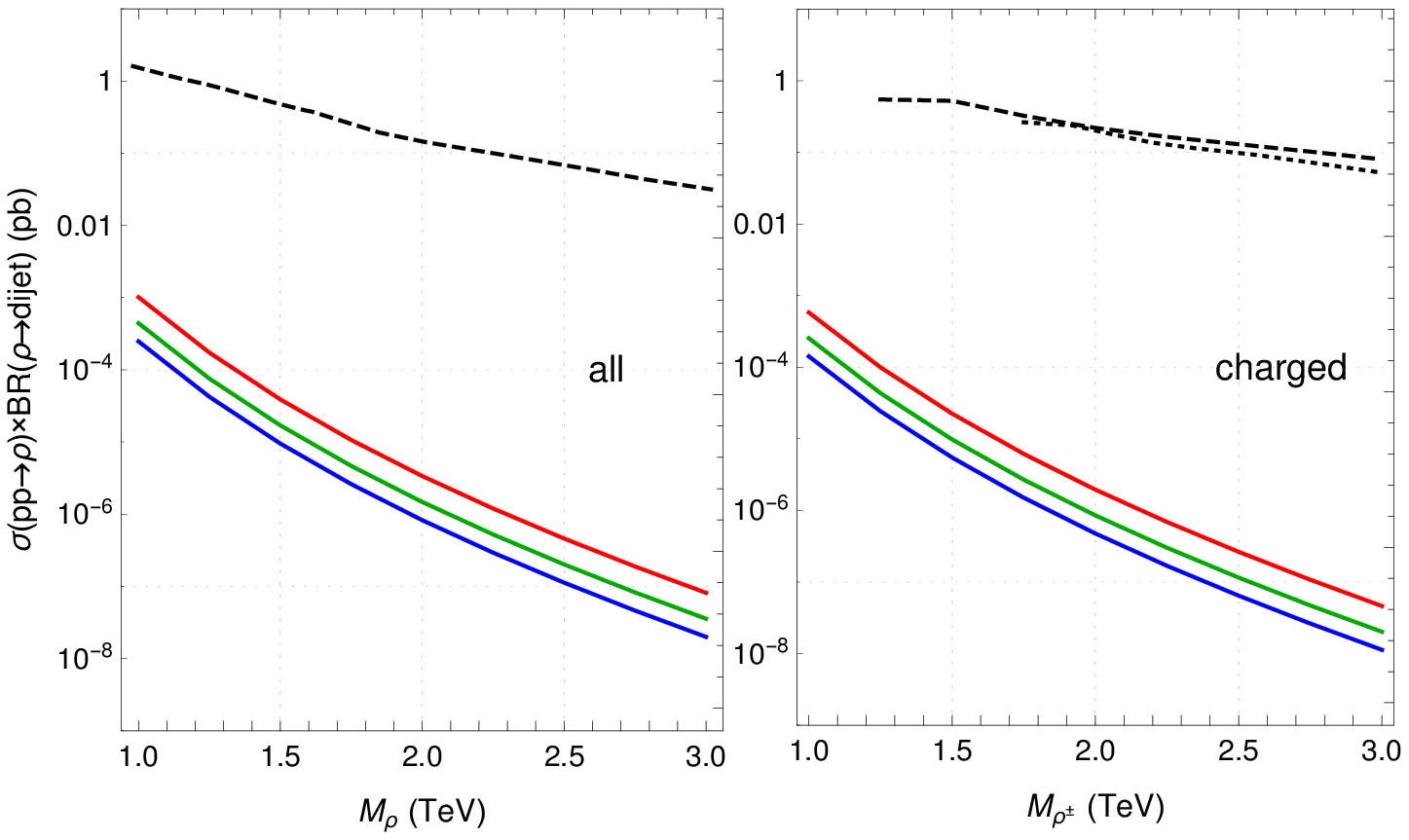}}
\caption{The $13\unit{TeV}$ cross section times branching ratios of our model
for the dijet (left panel) and charged dijet (right panel) decay
channels at $g''=10$ (solid upper red), $15$ (solid middle green), and $20$ (solid bottom blue).
In the left panel, the displayed experimental upper bound bellow $1.6\unit{TeV}$ is based on
$27\unit{fb}^{-1}$ of data and the bound above $1.6\unit{TeV}$ is based on $36\unit{fb}^{-1}$
of data~\cite{ExpXS-2016-CP056} (dashed).
In the right panel, the experimental bounds are based on $37\unit{fb}^{-1}$ of 
data~\cite{ExpXS-2017-EP-2017-042} (dotted), and
on $15.7\unit{fb}^{-1}$ of data~\cite{ExpXS-2016-AC069} (dashed).}
\label{Fig:XSxBRjets}
\end{figure}
In all these remaining channels, the predicted cross sections are too low when
compared to even most recent experimental upper bounds. 
This can be seen in Figs.~\ref{Fig:XSxBRVH}, \ref{Fig:XSxBRleptons}, and \ref{Fig:XSxBRjets}.
Thus there are no exclusion limits
implied by these channels. The introduction of the third quark family direct coupling 
will decrease the individual branching ratios of these channels by the same factors
as in the case of the $WW/WZ$ channels.
Therefore, no mass exclusion limits can emerge with this modification.

\section{Conclusions}
\label{sec:conclusions}

We have studied the production cross section times 
branching ratio for various decay channels in the productions of hypothetical neutral and 
charged vector resonances of new strong physics origin.
The resonances have been introduced in the context of 
the phenomenological Lagrangian where beside the composite $125\unit{GeV}$
Higgs boson the $SU(2)_{L+R}$ triplet of composite vector resonances is 
explicitly present. The ESB sector of our effective Lagrangian has been based on the 
$SU(2)_L\times SU(2)_R\rightarrow SU(2)_{L+R}$ non-linear 
sigma model while the scalar resonance has been introduced 
as the $SU(2)_{L+R}$ singlet. The vector resonance has been
built in employing the hidden local symmetry approach.
While allowed by the symmetry of the Lagrangian no direct fermion interactions
of the vector resonance triplet have been considered. The only interactions
of the resonance to fermions have been those generated by the mixing with
the EW gauge bosons.

We have compared the model's predictions with the upper bounds on the
production cross section times branching ratio obtained by the ATLAS and CMS 
Collaborations from the full 2016 data sets.
We have found that the $WW$ ($WZ$) channel excludes the mass of the neutral 
(charged) vector resonance
below about $2.1\unit{TeV}$ ($2.5\unit{TeV}$) and 
$1.1\unit{TeV}$ ($1.3\unit{TeV}$) for $g''=10$ and $15$, respectively.
The $g''=20$ cases are not restricted above $1\unit{TeV}$.
The direct fermionic interactions of the vector resonance can noticeably decrease 
the cross section predictions (and, thus, release the experimental restrictions) of 
the model in these channels.

The predicted cross sections are well below the experimental upper bounds
in the $bb$, $tt$, and $tb$ channels, as well as in all other channels considered
in this paper. The $bb$, $tt$, and $tb$ channels would be affected most
once the direct fermionic interactions of the vector resonance are introduced.
Consequently, the predicted cross sections in these channels might increase up 
to three orders of magnitude. The cross sections in all remaining channels
would shrink by the same factor as in the $WW$ and $WZ$ channels.

In summary, the new strong physics vector resonances of the considered type are
restricted by the current LHC data significantly weaker than their weakly-interacting
counter-parts. The lower exclusion limits will be further relaxed if there are
direct interactions of the vector resonances to the third generation quark doublet.

\section*{Acknowledgments}
We would like to thank Karol Kova\v{r}\'{i}k for useful discussions. The work of M.G. and J.J.
was supported by the Grants LTT17018 and LG15052 of the Ministry of Education, Youth
and Sports of the Czech Republic. M.G. was supported by the Slovak CERN Fund. J.J. was
supported by the National Scholarship Programme of the Slovak Republic.




\begin{thebibliography}{99}

\bibitem{125GeVBosonDiscovery}
G.~Aad \textit{et al.} (ATLAS Collaboration), Phys. Lett. B
\textbf{716}, 1 (2012); S.~Chatrchyan \textit{et al.} (CMS
Collaboration), \textit{ibid.} 30 (2012).

\bibitem{LeeQuiggThacker77}
B.~W.~Lee, C.~Quigg, and H.~B.~Thacker, Phys. Rev. D \textbf{16},
1519 (1977); 
Phys. Rev. Lett. \textbf{38} (1977) 883.

\bibitem{Chanowitz85}
M.~S.~Chanowitz and M.~K. Gaillard, Nucl. Phys. \textbf{B261}, 379
(1985).

\bibitem{HLS}
M.~Bando, T.~Kugo, and K.~Yamawaki, Phys. Rep. \textbf{164}, 217
(1988).

\bibitem{3siteHiggslessModel}
R.S.~Chivukula et al, Phys. Rev. D
\textbf{74}, 075011 (2006).

\bibitem{tBESSprd11}
M.~Gintner, J.~Jur\'{a}\v{n}, and I.~Melo, Phys. Rev. D
\textbf{84}, 035013 (2011).

\bibitem{tBESSepjc13}
M.~Gintner, J.~Jur\'{a}\v{n}, Eur. Phys. J. C
\textbf{73}, 2577 (2013).

\bibitem{tBESSepjc16}
M.~Gintner, J.~Jur\'{a}\v{n}, Eur. Phys. J. C
\textbf{76}, 651 (2016), erratum,  Eur.Phys.J. 
C\textbf{77}, 6 (2017).

\bibitem{Comm2017}
M.~Gintner, J.~Jur\'{a}\v{n}, to be published in
Communications -- Scientific Letters of the University of Zilina.

\bibitem{Pappadopulo_etal14}
D. Pappadopulo, A. Thamm, R. Torre and A. Wulzer, JHEP 1409 (2014) 060.

\bibitem{Dawson1985EWA}
S. Dawson, Nucl.Phys. B 249, 42 (1985).

\bibitem{Mathematica10_2016}
Wolfram Research, Inc., Mathematica, Version 10.4, Champaign, IL (2016).

\bibitem{ManeParse2016}
D.B.~Clark, E.~Godat, F.I.~Olness, arXiv:1605.08012. 
Mane Parse package download: https://ncteq.hepforge.org/mma/index.html

\bibitem{HepForgeCT10_2015}
A.~Buckley et al., Eur. Phys. J. C 75, 132 (2015); arXiv:1412.7420. 
LHAPDF6 PDFs download: http://lhapdf.hepforge.org/pdfsets

\bibitem{ExpXS-N1-B2G-17-001}
CMS Collaboration, CMS PAS B2G-17-001.

\bibitem{ExpXS-2016-AC062}
ATLAS Collaboration, ATLAS-CONF-2016-062.

\bibitem{ExpXS-2016-AC082}
ATLAS Collaboration, ATLAS-CONF-2016-082.

\bibitem{ExpXS-2016-AC031}
ATLAS Collaboration, ATLAS-CONF-2016-031.

\bibitem{ExpXS-2016-AC060}
ATLAS Collaboration, ATLAS-CONF-2016-060.

\bibitem{ExpXS-2017-EP-2017-049}
CMS Collaboration, arXiv:1704.03366.

\bibitem{ExpXS-2016-AC014}
ATLAS Collaboration, ATLAS-CONF-2016-014.

\bibitem{ExpXS-N3-B2G-17-010}
CMS Collaboration, CMS PAS B2G-17-010

\bibitem{ExpXS-2017-AC018}
ATLAS Collaboration, ATLAS-CONF-2017-018.

\bibitem{ExpXS-2017-CP002}
CMS Collaboration, CMS PAS B2G-17-002.

\bibitem{ExpXS-2016-AarXiv2}
ATLAS Collaboration, arXiv:1607.05621.

\bibitem{ExpXS-2017-AC027}
ATLAS Collaboration, ATLAS-CONF-2017-027.

\bibitem{ExpXS-2017-AC016}
ATLAS Collaboration, ATLAS-CONF-2017-016.

\bibitem{ExpXS-2017-CMS-JHEP02}
CMS Collaboration, JHEP 02(2017)048.

\bibitem{ExpXS-2016-CP056}
CMS Collaboration, CMS PAS EXO-16-056.

\bibitem{ExpXS-2017-EP-2017-042}
ATLAS Collaboration, arXiv:1703.09127.

\bibitem{ExpXS-2016-AC069}
ATLAS Collaboration, ATLAS-CONF-2016-069.

\end{thebibliography}
\end{document}